\documentclass[aps,pra,superscriptaddress,amsmath,amssymb,preprintnumbers,twocolumn]{revtex4}
\usepackage{amssymb}
\usepackage{booktabs}
\usepackage{graphicx}
\usepackage{bm}
\usepackage{color}
\usepackage{cancel}
\usepackage{slashed}
\usepackage{subfigure}

\newcommand{\Ignore}[1]{}

\newcommand{\Ket}[1]{\left\vert #1\right\rangle}
\newcommand{\Bra}[1]{\left\langle #1\right\vert}

\newcommand{\ii}{\mathrm{i}}
\newcommand{\ee}{\mathrm{e}}

\begin{document}

\title{Detuning-induced robustness of a three-state Lanzau-Zener model against dissipation}

\author{Benedetto Militello}
\address{Universit\`a degli Studi di Palermo, Dipartimento di Fisica e Chimica - Emilio Segr\`e, Via Archirafi 36, I-90123 Palermo, Italy}
\address{I.N.F.N. Sezione di Catania, Via Santa Sofia 64, I-95123 Catania, Italy}

\begin{abstract}

A three-state system subjected to a time-dependent Hamiltonian whose bare energies undergo one or more crossings, depending on the relevant parameters, is considered, also taking into account the role of dissipation in the adiabatic following of the Hamiltonian eigenstates.
Depending on the fact that the bare energies are equidistant or not, the relevant population transfer turns out to be very sensitive to the environmental interaction or relatively robust. The physical mechanisms on the basis of this behavior are discussed in detail.
 
\end{abstract}

\maketitle

\section{Introduction}\label{sec:introduction}

The model of Landau-Zener-Majorana-Stueckelberg (LZMS)~\cite{ref:Landau,ref:Zener,ref:Majo,ref:Stuck} is an important solvable model with a time-dependent Hamiltonian. 
Dynamical problems for non stationary systems are usually hard to solve unless particular conditions are satisfied~\cite{ref:Barnes2012,ref:Simeonov2014,ref:Sinitsyn2018,ref:Sriram2005,ref:Owusu2010,ref:Chrusc2015}. For very slowly changing Hamiltonians the adiabatic approximation is applicable~\cite{ref:Messiah,ref:Griffiths}, and, in the case of LZSM model, deviations from adiabaticity can be evaluated through the remarkable formula derived in Refs~\cite{ref:Landau,ref:Zener,ref:Majo,ref:Stuck}.
The LZMS model describes a two-state system with linearly time-dependent bare (diabatic) energies and a stationary coupling between the bare eigenstates, which is responsible for avoiding the crossing that would occur at some instant of time, leading to dressed (adiabatic) energies which remain well separated. In the original mathematical formulation the experiment is assumed to have an infinite duration. The implications of the inevitably finite duration of a real experiment have been considered~\cite{ref:Vitanov1996,ref:Vitanov1999a}. The hypothesis of linear time-dependence of the bare energies has also been relaxed~\cite{ref:Vitanov1999b}. Systems governed by nonlinear equations~\cite{ref:Ishkhanyan2004,ref:MilitelloPRE2018} and non-Hermitian Hamiltonian models~\cite{ref:Toro2017} have been considered.
The case where neither the bare energies nor the dressed ones cross (the hidden crossing model) has been deeply studied~\cite{ref:Fishman1990,ref:Bouwm1995}, as well as the opposite situation (the total crossing model), where, because of a suitable time-dependence of the coupling constant, both the bare and the dressed energies cross~\cite{ref:Militello2015a}.

In his seminal work~\cite{ref:Majo}, Majorana considered a spin-$j$ immersed in a magnetic field with linearly changing $z$-component, therefore studying the dynamics of a system in the presence of a multi-level crossing. In this scheme, due to the presence of a static transverse component of the field, each level is coupled to the immediately upper one.
Other important multi-level crossing models are the so called equal-slope model and the bow-tie model. In the first one, introduced by Demkov and Osherov~\cite{ref:Demkov1967}, a single time-varying level intercepts a series of static energies, producing a sequence of two-state crossings, which allows to study the relevant dynamics through the so called Independent Crossing Approximation~\cite{ref:Brundobler1993}. 
On the contrary, the bow-tie model consists of $N$ states having bare energies which cross at the same time, hence realizing a proper multi-state crossing. Moreover, in this model one state is coupled to the remaining $N-1$, which on the other hand do not couple each other. The $N=3$ situation has been introduced and studied in depth by  Carroll and Hioe~\cite{ref:Carroll1986a,ref:Carroll1986b}, but it is the case to observe that the Majorana model for a spin-$1$ is a special case of $N=3$ bow-tie model. The scenario with $N-2$ decoupled states crossing at the same time and two states interacting with the remaining $N-2$ has also been considered as a possible generalization of the Carroll-Hioe model~\cite{ref:Demkov2000}. 
The degenerate Landau-Zener model, consisting of two degenerate levels which cross~\cite{ref:Vasilev2007}, and a hybrid model between LZSM and Stimulated Raman Adiabatic Passage with time-dependent coupling constants~\cite{ref:Ivanov2008} have also been introduced. Other  specific models with a fixed number of states and particular coupling configurations have been considered~\cite{ref:Shytov2004,ref:Sin2015,ref:Li2017,ref:Sin2017}.
Time-dependent Rabi Hamiltonian~\cite{ref:Dodo2016} and Tavis-Cummings model~\cite{ref:Sun2016,ref:Sin2016} exhibit dynamical behaviors that can be traced back to multi-level LZMS transitions.
 
There are several studies on the effects of dissipation and decoherence on adiabatic evolutions in general~\cite{ref:Lidar,ref:Florio,ref:ScalaOpts2011,ref:Wild2016} and, in partitular, on two-state LZMS processes~\cite{ref:Ao1991,ref:Potro2007,ref:Wubs2006,ref:Saito2007,ref:Lacour2007,ref:Nel2009,ref:ScalaPRA2011}. In spite of this, environmental effects on adiabatic evolutions in the presence of crossings involving more than two states have been studied only rarely. Quite recently, Ashhab~\cite{ref:Ashhab2016} has analyzed the multi-level LZMS problem in the presence of an interaction with the environment, focusing on dephasing. More recently, a dissipative three-state LZMS problem has been considered~\cite{ref:Militello2019}.

In this paper we consider a three-state LZMS similar to the one considered in Ref.~\cite{ref:Militello2019}, with two time-dependent bare energies, a time-independent one and a cyclic coupling scheme. Differently from the model previously considered, the bare energies are not equidistant, since the middle level presents a static energy offeset, i.e. a nonzero bare energy, which determines the occurrence of a series of binary crossings. The system is considered in the absence and in the presence of environmental interaction, bringing to light the positive role of the detuning in a certain range of values. 
Experiments have been developed in the context of artificial atoms where a multi-state system undergoes several independent crossings in the presence of environment-induced effects~\cite{ref:Berns2008}. Though our model does not perfectly match that situation, our analysis constitutes an advancement in the comprehension of the role of environmental interaction in multi-state Landau-Zener processes.
In the next section, we introduce the Hamiltonian model and analyze the population transfer process in the absence of any dissipation, singling out the role of the detuning in the way the level crossings occur. In sec.~\ref{sec:dissipative} we add the role of the environment, analyzing the competition between the detrimental action of the decays and the positive effect of the detuning. Finally, in sec.~\ref{sec:conclusions} we give some conclusive remarks.

\section{Ideal system}\label{sec:ideal}

{\it The model --- } We consider a three-state system with a time-independent cyclic coupling scheme and two linearly time-dependent bare energies ($\hbar=1$):
\begin{eqnarray}\label{eq:Hamiltonian_ideal}
H(t) = \left(
\begin{array}{ccc}
-\kappa t & \Omega_{12} & \Omega_{13} \\
\Omega_{12} & \Delta & \Omega_{23} \\
\Omega_{13} & \Omega_{23} & \kappa t 
\end{array}
\right)\,, \label{eq:IdealHam}
\end{eqnarray}
where, for the sake of simplicity, only real coupling strengths are considered. We will refer to the energy offset $\Delta$ 
as detuning, independently from its physical origin.
This model can be easily obtained in superconducting qutrits subjected to cyclic couplings~\cite{ref:NoriPRL2005,ref:NoriSR2014}. Moreover, it can describe the dynamics in a two-qubit triplet subspace in the presence of an external magnetic field (with a linearly time-dependent $z$-component) and an exchange interaction~\cite{ref:Heisenberg1,ref:Heisenberg2,ref:Heisenberg3}. In this second scenario, in addition to the interaction terms described in Ref.~\cite{ref:Militello2019} where the $\Omega_{12}=\Omega_{23}$ has been studied, a longitudinal interaction has to be considered, leading to $H=\kappa t (\sigma_z^\mathrm{A}+\sigma_z^\mathrm{B}) + \Omega_{12} (\sigma_x^\mathrm{A}+\sigma_x^\mathrm{B}) + \Omega_{13}/2 (\sigma_x^\mathrm{A}\sigma_x^\mathrm{B} - \sigma_y^\mathrm{A}\sigma_y^\mathrm{B}) -\Delta/2 \sigma_z^\mathrm{A}\sigma_z^\mathrm{B}$, whose restriction in the invariant subspace $\{\Ket{\uparrow\uparrow}, (\Ket{\uparrow\downarrow}+\Ket{\downarrow\uparrow})/\sqrt{2}, \Ket{\downarrow\downarrow}\} \equiv \{\Ket{1}, \Ket{2}, \Ket{3}\}$ coincides with the Hamiltonian in \eqref{eq:Hamiltonian_ideal} up to a global shift. 

When $t$ spans the time interval $[-T, T]$, with $T$ very large, we have that the highest instantaneous eigenvalue of the Hamiltonian (almost perfectly) corresponds to $\Ket{1}$ for $t=-T$ and (almost perfectly) to $\Ket{3}$ for $t=T$. Therefore, if the adiabatic approximation is valid in the whole time interval, the adiabatic following of this eigenstate determines a complete $\Ket{1}\rightarrow\Ket{3}$ population transfer.

\begin{figure}[h]
\includegraphics[width=0.45\textwidth, angle=0]{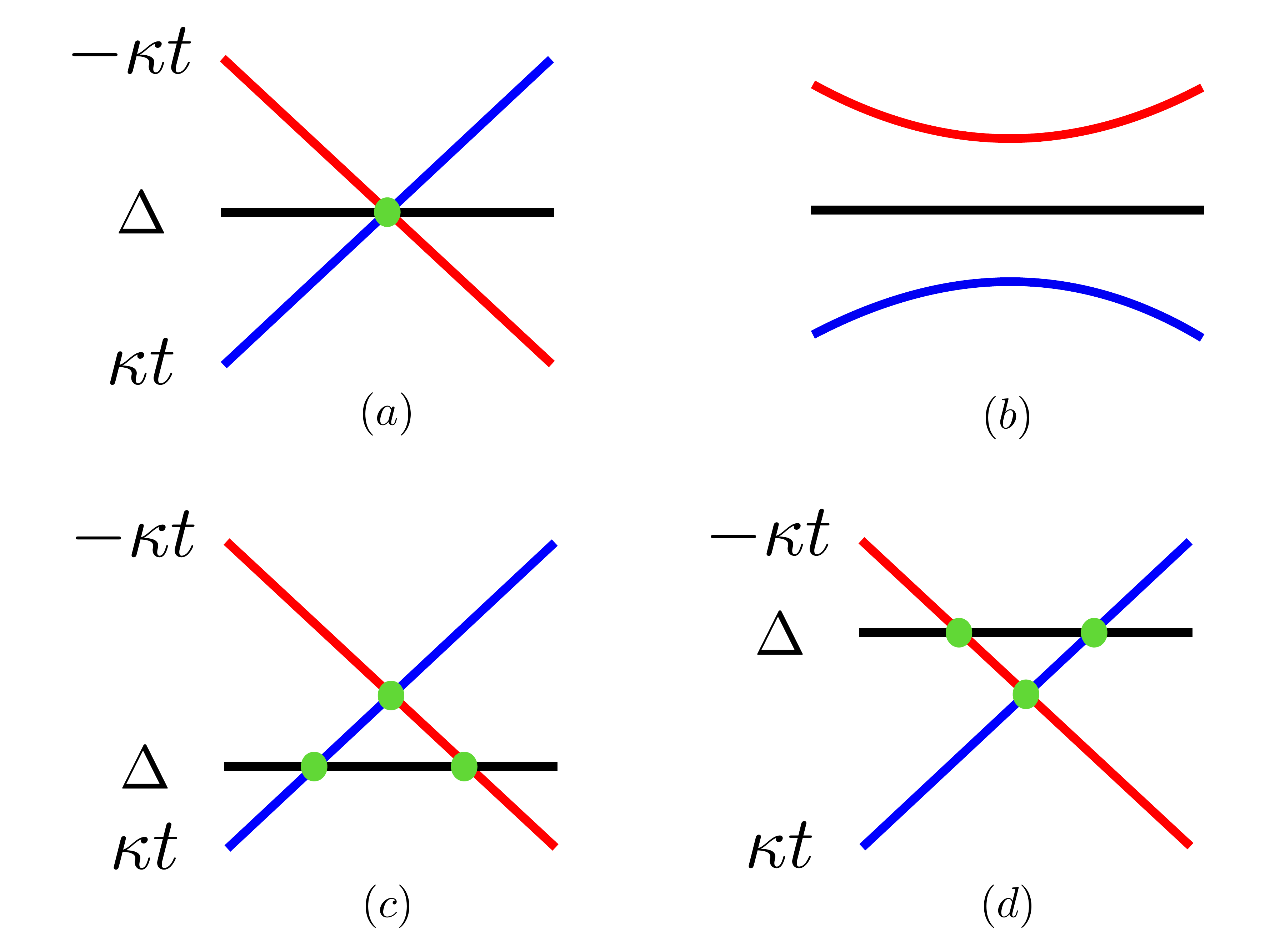} 
\caption{(Color online) Level crossing schemes for different values of $\Delta$:  the three bare levels for $\Delta=0$ (a) and the corresponding dressed energies when $\Omega_{12}=\Omega_{23}$ and $\Omega_{13}=0$ (b); the bare energies for $\Delta<0$ (c) and $\Delta>0$ (d).} \label{fig:levelscheme}
\end{figure}

{\it Independent Crossings Approximation --- } The detuning contributes to determine the pattern of the level crossings occurring in the time interval $[-T,T]$, which in turn can affect the population transfer. In fact, in the $\Delta=0$ case the three bare (diabatic) energies simultaneously touch and cross at $t=0$. On the contrary, a non vanishing detuning makes the bare energies cross in pairs at three different times. For example, for $\Delta<0$ state $\Ket{3}$ with bare energy $\kappa t$ intercept the second bare level $\Delta$ at $t=\Delta/\kappa$, then bare energies of states $\Ket{1}$ and $\Ket{3}$ cross at $t=0$ and, finally, bare energies of states $\Ket{1}$ and $\Ket{2}$ cross for $t=-\Delta/\kappa$.  For $\Delta>0$ the bare energy crossings occur in the reversed order. The three situations are illustrated in Figs.~\ref{fig:levelscheme}a, \ref{fig:levelscheme}c and \ref{fig:levelscheme}d. In Fig.~\ref{fig:levelscheme}b is shown an example of dressed (adiabatic) energies for $\Delta=0$.

The way the bare energy crossings occur and the couplings between the relevant states influence the efficiency of the population transfer. In Fig.~\ref{fig:Pop3Ideal-1}a is shown the population of the state $\Ket{3}$ at $t=T$ as a function of $\Delta$, for different values of $\Omega_{12}$, assuming $\Omega_{13}=0$ and $\Ket{1}$ as the initial state. The population transfer turns out to be very efficient for $\Delta>0$, while for $\Delta<0$ the efficiency becomes lower and lower as $\Omega_{12}$ assumes smaller values. This behavior can be well understood in terms of the crossing pattern. In fact, for $\Delta<0$ bare energies of states $\Ket{3}$ and $\Ket{2}$ cross first, and nothing happens since all the population is present in state $\Ket{1}$, at this stage. Then, at $t=0$, bare energies of states $\Ket{1}$ and $\Ket{3}$ cross and a certain amount of population is transferred from $\Ket{1}$ to $\Ket{3}$. Whether the transfer is complete or not, it depends on the coupling strength $\Omega_{13}$, which is zero, in our case. Finally, the crossing between $\Ket{1}$ and $\Ket{2}$ occurs, with no influence on the population of state $\Ket{3}$. 
For $\Delta>0$ the $\Ket{1}$-$\Ket{2}$ crossing occurs first, bringing population to state $\Ket{2}$, provided $\Omega_{12}$ is large enough. The subsequent $\Ket{1}$-$\Ket{3}$ crossing is irrelevant, if all the population has been transferred to state $\Ket{2}$; moreover, in our case $\Omega_{13}=0$, so that nothing happens at this stage even if the transfer during the first crossing was imperfect. Finally, in the last crossing involving the $\Ket{2}$-$\Ket{3}$ states the population previously transferred to $\Ket{2}$ moves toward $\Ket{3}$. Of course, if the coupling strengths $\Omega_{12}$ and $\Omega_{23}$ are not large enough the two population transfers are incomplete or even absent. This is why for small or vanishing values of $\Omega_{12}$ the final population of $\Ket{3}$ is small even for positively large $\Delta$, which is well visible in  Fig.~\ref{fig:Pop3Ideal-1}.    

In Fig.~\ref{fig:Pop3Ideal-1}c the case where $\Omega_{13}/\Omega_{23}=0.5$ is considered, which implies that during the $\Ket{1}$-$\Ket{3}$ crossing a population transfer can occur.  For this reason, when $\Delta<0$ the total process is very efficient. In fact, the first and third crossings do not have significant implications, since the involved states are not populated when they occur: only the state $\Ket{1}$ is populated when $\Ket{2}$ and $\Ket{3}$ cross, and, provided a complete transfer has occurred during the $\Ket{1}$-$\Ket{3}$ crossing, the interception of bare energies of states $\Ket{1}$ and $\Ket{2}$ is irrelevant as well. The whole process is efficient as much as the population transfer during the second crossing is complete. For $\Delta>0$, assuming $\Omega_{12}$ large enough, at the first crossing all the population is transferred to the state $\Ket{2}$, then nothing happens during the second crossing involving $\Ket{1}$ and $\Ket{3}$, since both states are \lq empty\rq. Finally, during the third crossing, the population is transferred from $\Ket{2}$ to $\Ket{3}$, provided $\Omega_{23}$ is adequate. If $\Omega_{12}$ is too small, the transfer in the first crossing is compromised, and so is the entire process.

{\it Adiabatic eigenstates --- } The treatment based on the independent crossing approximation turns out to be good enough for (negatively or positively) large $\Delta$, since in such a case the three crossings are well separated. On the contrary, when $\Delta$ is of the same order of the coupling strengths, the proper way to analyze the system is to consider the dressed (adiabatic) energies. Of course, this approach is valid in every regime.

In Figs.~\ref{fig:Pop3Ideal-1}b and \ref{fig:Pop3Ideal-1}d is reported the minimum energy gap between the two highest eigenvalues of the Hamiltonian in the time interval $[-T,T]$. More precisely, given a set of parameters which define $H(t)$, we have three functions $\epsilon_1(t)\ge \epsilon_2(t)\ge \epsilon_3(t)$ that correspond to the three instantaneous eigenvalues of the Hamiltonian; consider that $\epsilon_1(\pm T)\approx \kappa T$ and that the corresponding eigenstate, satisfying $\Ket{\epsilon_1(-T)}\approx \Ket{1}$ and $\Ket{\epsilon_1(T)}\approx \Ket{3}$, is the one which is expected to carry population from $\Ket{1}$ to $\Ket{3}$. On this basis we define:
\begin{equation}\label{eq:mingap}
{\cal G} = \mathrm{min}_{t\in [-T,T]}\left\{\epsilon_1(t)-\epsilon_2(t)\right\}\,.
\end{equation}

According to the general theory of the adiabatic approximation~\cite{ref:Messiah,ref:Griffiths}, diabatic transitions can occur when the square of an energy gap turns out to be not much larger than the relevant matrix element of $\dot{H}$, which in our case linearly depend on $\kappa$. Therefore, if $\cal G$ is smaller then or comparable to $\sqrt{\kappa}$, then the adiabatic approximation fails at some instant of time, which jeopardizes the transportation of population from $\Ket{1}$ to $\Ket{3}$ through the eigenstate corresponding to $\epsilon_1(t)$.
On the contrary, a high value of $\cal G$ guarantees the validity of the adiabatic approximation in the whole time interval and, as a consequence, a complete population transfer.
Fig.~\ref{fig:Pop3Ideal-1} clearly illustrates this connection.

\begin{widetext}

\begin{figure}[h]
\begin{tabular}{cc}
\begin{tabular}{cc}
\subfigure[]{\includegraphics[width=0.32\textwidth, angle=0]{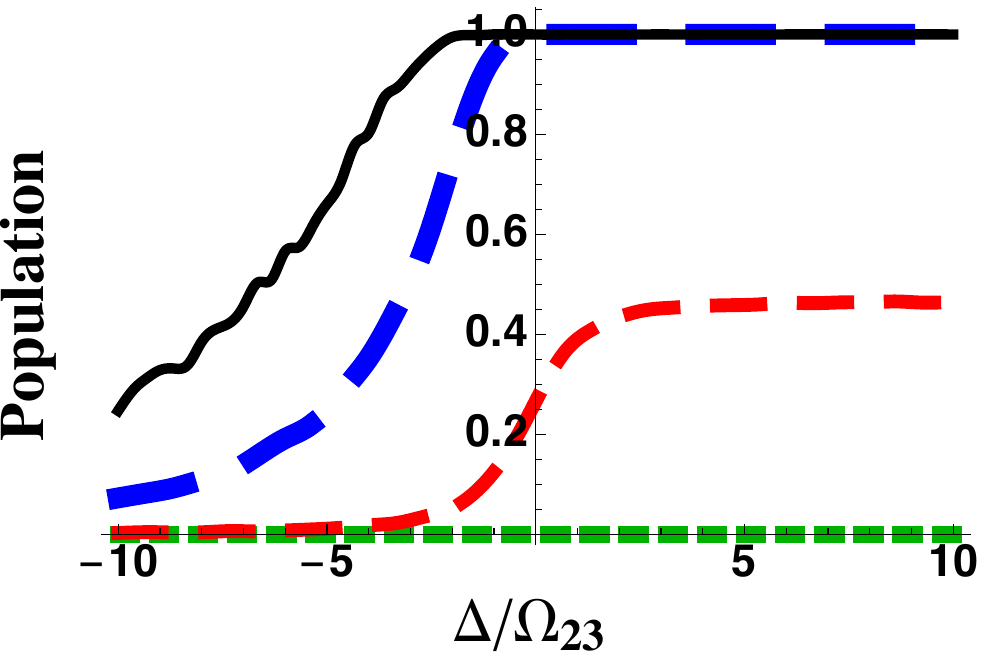}} &
\hskip1cm
\subfigure[]{\includegraphics[width=0.32\textwidth, angle=0]{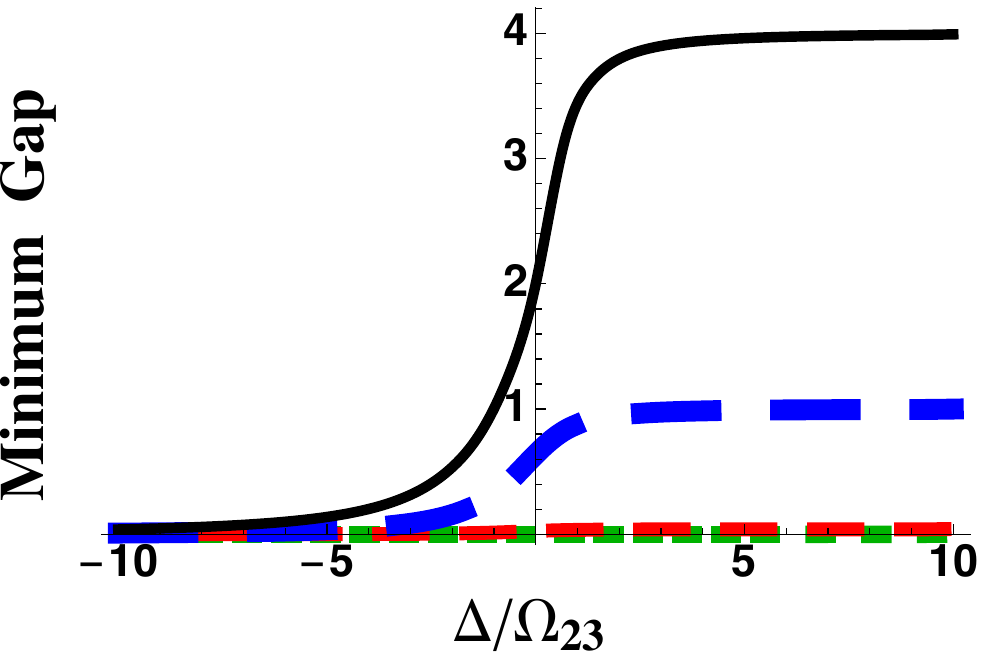}} \\
\subfigure[]{\includegraphics[width=0.32\textwidth, angle=0]{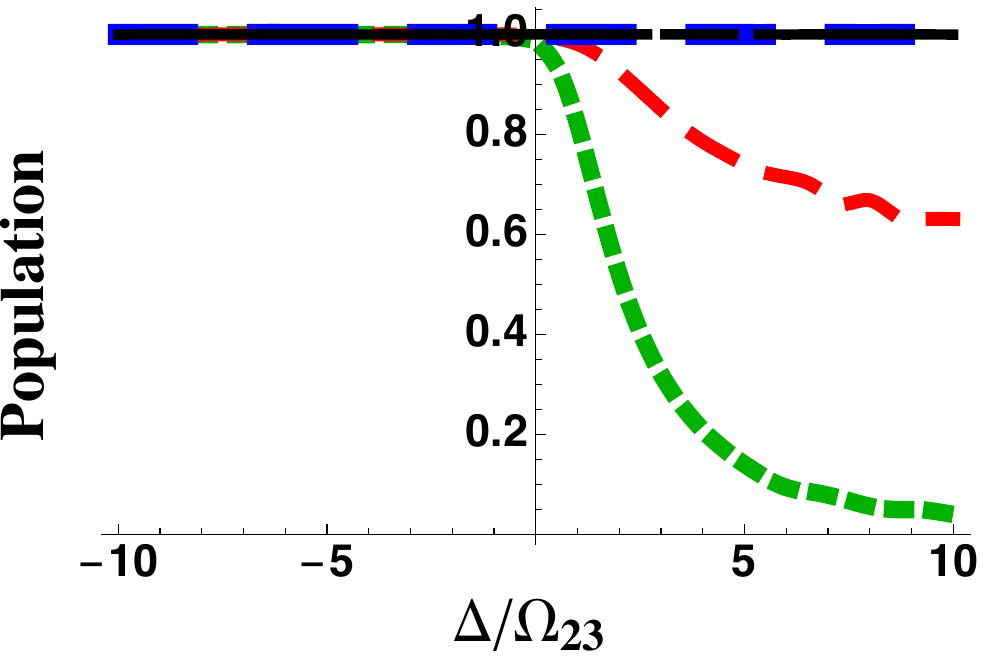}} &
\hskip1cm
\subfigure[]{\includegraphics[width=0.32\textwidth, angle=0]{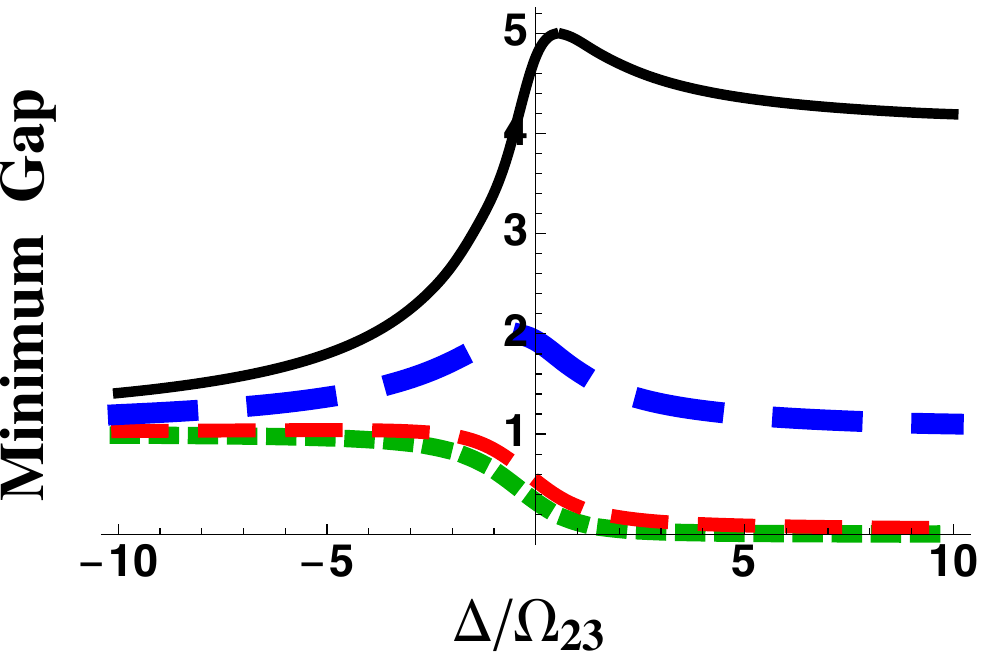}} 
\end{tabular} & \hskip0.7cm
\begin{tabular}{c}
\includegraphics[width=0.14\textwidth, angle=0]{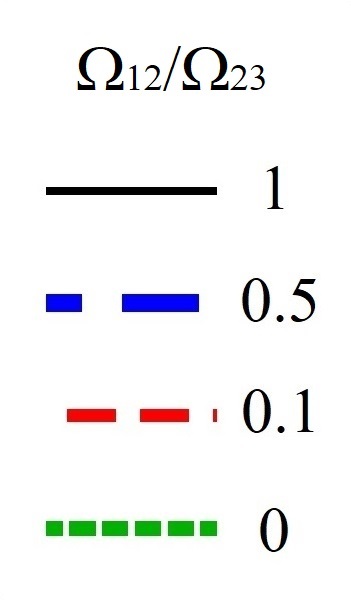}
\end{tabular}
\end{tabular}
\caption{(Color online) Final population of state $\Ket{3}$ (obtained through a numerically exact resolution of the relevant Schr\"odinger equation) when the system starts with $\Ket{1}$ as a function of $\Delta$ (in units of $\Omega_{23}$), and relevant minimum energy gap between the two highest instantaneous eigenvalues of the Hamiltonian  (in units of $\Omega_{23}$). 
In (a) and (b) the case $\Omega_{13}/\Omega_{23}=0$ is considered, while in (c) and (d) we have $\Omega_{13}/\Omega_{23}=0.5$.
The other parameters are: $\kappa/\Omega^2_{23}=0.1$ and $\kappa T / \Omega_{23} = 100$ for all the curves. 
In all figures, the four curves correspond to: $\Omega_{12}/\Omega_{23}=1$ (solid black line), $\Omega_{12}/\Omega_{23}=0.5$ (bold long dashed blue line), $\Omega_{12}/\Omega_{23}=0.1$ (dashed red line), $\Omega_{12}/\Omega_{23}=0$ (dotted green line).
}\label{fig:Pop3Ideal-1}
\end{figure}

\begin{figure}[h]
\begin{tabular}{cclrr}
\subfigure[]{\includegraphics[width=0.27\textwidth, angle=0]{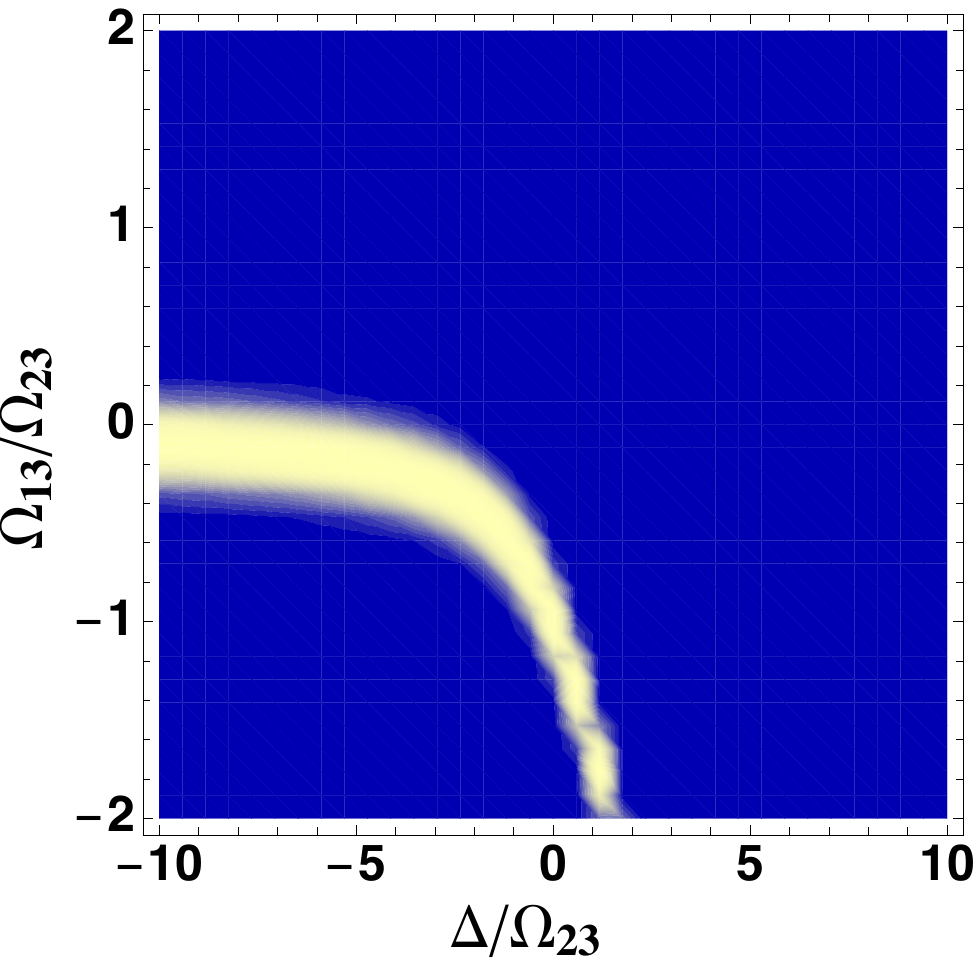}} &
\subfigure[]{\includegraphics[width=0.27\textwidth, angle=0]{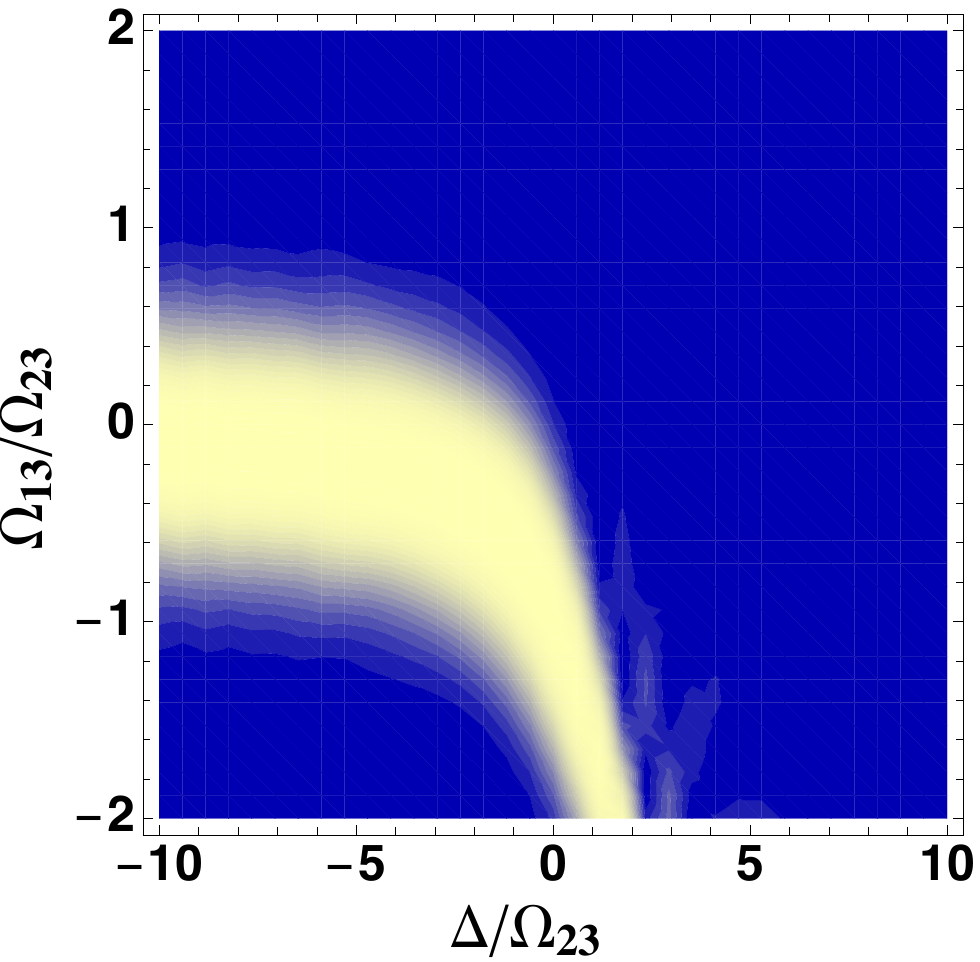}} &
\includegraphics[width=0.06\textwidth, angle=0]{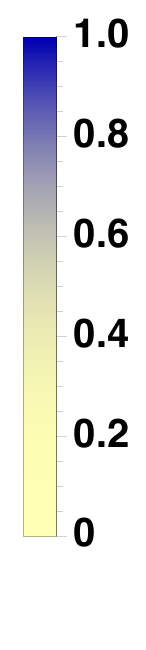} &
\subfigure[]{\includegraphics[width=0.27\textwidth, angle=0]{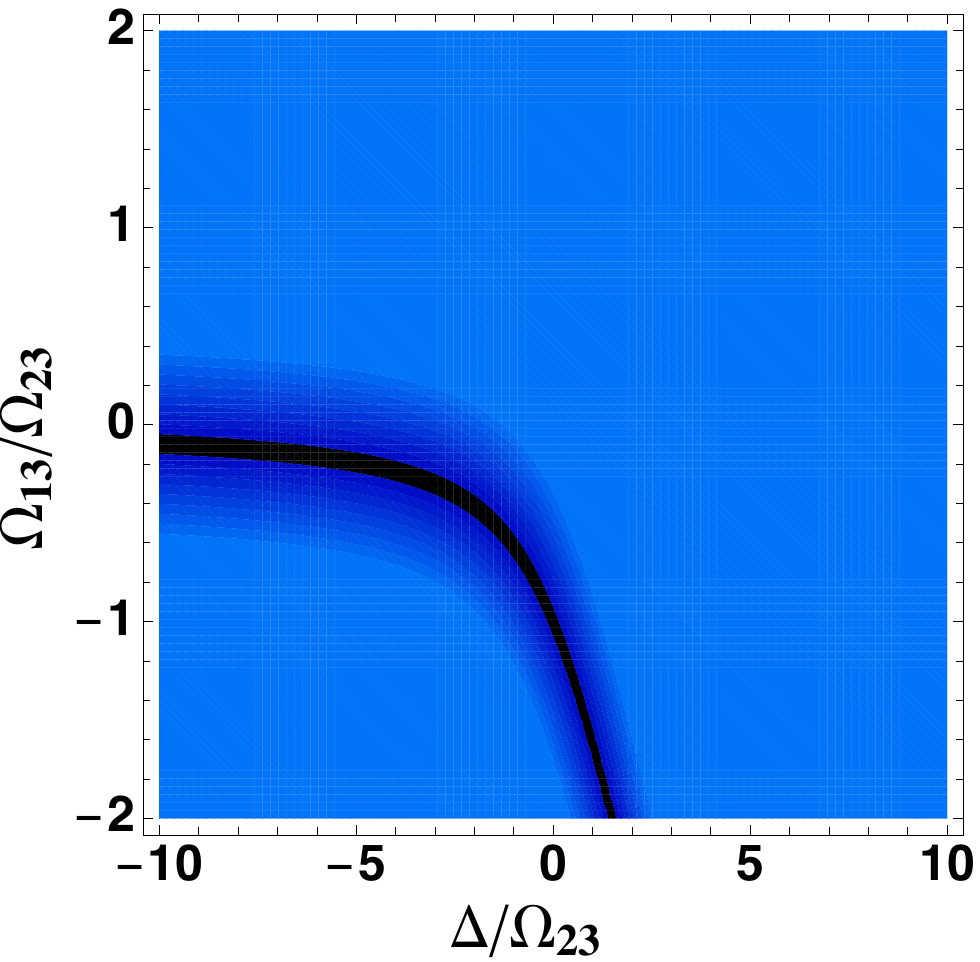}} &
\subfigure{\includegraphics[width=0.06\textwidth, angle=0]{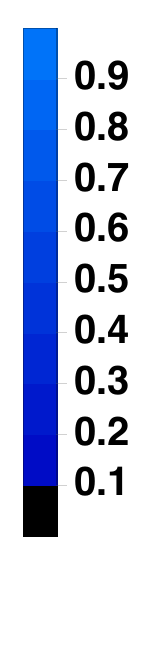}} 
\end{tabular}
\caption{(Color online) Final population of state $\Ket{3}$ (obtained through a numerically exact resolution of the relevant Schr\"odinger equation) when the system starts in $\Ket{1}$ as a function of $\Omega_{13}$ and $\Delta$ (both in units of $\Omega_{23}$). In (a) $\kappa/\Omega_{23}^2=0.1$ while in (b)  $\kappa/\Omega_{23}^2=1$; in both cases $\Omega_{12}=\Omega_{23}$ and $\kappa T/\Omega_{23}=100$. In (c) it is represented the relevant minimum energy gap between the two highest instantaneous eigenvalues of the Hamiltonian as a function of $\Omega_{13}$ and $\Delta$ (all in units of $\Omega_{23}$).
}\label{fig:Pop3Ideal-2}
\end{figure}

\end{widetext}

In Fig.~\ref{fig:Pop3Ideal-2} is shown the efficiency of the population transfer as a function of the $\Ket{1}$-$\Ket{3}$ coupling strength and the detuning, for two different values of the changing rate of the bare energies $\kappa$. In Fig.~\ref{fig:Pop3Ideal-2}c is also shown the behavior of $\cal G$ in the same parameter region. (It is the case to observe that the value of the minimum gap does not depend on the value of $\kappa$ when $t$ goes from $-\infty$ to $+\infty$, or when the time interval $[-T,T]$ is $\kappa$-dependent in such a way that  $\kappa T$ is always the same, which is our case for Figs.~\ref{fig:Pop3Ideal-2}a and \ref{fig:Pop3Ideal-2}b). It is well visible that in the zone where the minimum energy gap becomes very small the efficiency gets quite low both for $\kappa/\Omega_{23}^2=0.1$ and $\kappa/\Omega_{23}^2=1$. Moreover, since we have to compare the gap with $\sqrt{\kappa}$, for larger values of this parameter the zone of low efficiency is wider.

{\it Role of possible phases --- } At first glance, the model in \eqref{eq:IdealHam} can be considered as a generalization of the model in Ref.~\cite{ref:Militello2019}, because of the nonzero detuning and the independence of the three coupling strengths $\Omega_{ij}$'s (in Ref.~\cite{ref:Militello2019} we have $\Omega_{12}=\Omega_{23}$). Nevertheless, the latter model includes the possibility of complex coupling strengths, which are not considered in the former. Therefore, the two models are to be considered as different, none of them being the generalization of the other. It is anyway interesting to shortly comment on what happens if we include the phases. In fact, by the replacement $\Omega_{ij} \rightarrow \Omega_{ij}\, \mathrm{e}^{-\ii\phi_{ij}}$, the secular equation to determine the eigenvalues of the Hamiltonian in \eqref{eq:IdealHam} assumes the form: $\lambda^3-\Delta\lambda^2-(\Omega_{12}^2+\Omega_{23}^2+\Omega_{13}^2+\kappa^2 t^2) \lambda - 2 \Omega_{12}\Omega_{23}\Omega_{13}\cos(\phi_{12}+\phi_{23}-\phi_{13}) + (\Omega_{12}^2-\Omega_{23}^2)\kappa t + (\kappa^2 t^2 + \Omega_{13}^2) \Delta$. We immediately note that, in spite of having three independent phases, they appear in the equation only once and as a precise combination. Moreover, several calculations of the parameter ${\cal G}$ spanning the relevant parameters have shown that the degeneration jeopardizing the population transfer occurs only when $\phi_{12}+\phi_{23}-\phi_{13}\approx \pi$. This means that, for our purposes, considering the three coupling strength as complex numbers with independent phases or as real numbers spanning positive and negative values is essentially the same. Therefore, for the sake of simplicity we have omitted the complex phases.

\section{Role of Dissipation}\label{sec:dissipative}

Since the environment plays a certain role in experiments involving Landau-Zener transitions~\cite{ref:Zhou2014,ref:Sun2015,ref:Wang2016}, even multi-state ones~\cite{ref:Berns2008}, we want to analyze the effects of dissipation and decoherence on the population transfer.
Following the same approach of Ref.~\cite{ref:Militello2019}, we consider external decays, i.e. decays toward states which are orthogonal to the \lq main\rq\, subspace (generated by $\Ket{1}$, $\Ket{2}$ and $\Ket{3}$) we are focusing on. In such a case, a possible way to describe the zero-temperature evolution of the system is by using an appropriate master equation where the bare states are incoherently coupled (through the environment) to the external states~\cite{ref:Petru,ref:Gardiner}: 
\begin{eqnarray}
\nonumber
&& \dot\rho = -\ii [H(t),\rho] + \\ 
&& \sum_{j=1}^3 \sum_k \gamma_{kj} \left(\Ket{E_k}\Bra{j}\rho\Ket{j}\Bra{E_k}-\frac{1}{2}\{\Ket{j}\Bra{j}, \rho\}\right) \,, 
\end{eqnarray}
where $\Ket{E_k}$ is the $k$-th external state, $\Ket{j}$'s refer to the main subspace and $\gamma_{ij}$ are the relevant decay rates. 
Restricting the master equation to the main subspace is equivalent to consider the dynamics induced by a non-Hermitian Hamiltonian (see for example Refs.~\cite{ref:MilitelloPRA2010,ref:MilitelloOSID2016,ref:Napoli2004}) where the diagonal terms have imaginary parts given by the decay rates: $-\ii\Gamma_j = -\ii \sum_k \gamma_{kj}$ to be added to the $j$-th diagonal term of $H(t)$. It is worth mentioning that we have also assumed the absence of a direct coherent coupling between every two states $\Ket{j}$ and $\Ket{E_k}$, otherwise, an additional term in the commutator would be required. The Hamiltonian we obtain is
\begin{eqnarray}
H_\mathrm{D}(t) = \left(
\begin{array}{ccc}
-\kappa t -\ii\Gamma_1 & \Omega_{12} &  \Omega_{13} \\
\Omega_{12} & \Delta -\ii\Gamma_2 & \Omega_{23} \\
 \Omega_{13} & \Omega_{23} & \kappa t  -\ii\Gamma_3
\end{array}
\right)\,.
\end{eqnarray}

Following the same reasoning of Ref.~\cite{ref:Militello2019}, the presence of decays for state $\Ket{1}$ or $\Ket{3}$ dramatically compromise the efficiency of the population transfer, because the adiabatic state $\Ket{\epsilon_1(t)}$ almost coincides with $\Ket{1}$ for a long time from $-T$ to a time close to $t=0$, and is close to $\Ket{3}$ in the mirror interval, from a time near $t=0$ to $T$. In both cases a significant loss of population is experienced.
We then focus on the nontrivial effects of a decaying state $\Ket{2}$, then always assuming in the following $\Gamma_1=\Gamma_3=0$ and renaming the second decay as $\Gamma\equiv\Gamma_2$.

In Fig.~\ref{fig:Dissipation-1} it is shown the efficiency of the population transfer in the same parameter region considered for Fig.~\ref{fig:Pop3Ideal-2}a, but in the presence of dissipation. 
Three different values of $\Gamma$ are taken into account: $\Gamma/\Omega_{23}=0.001$, $\Gamma/\Omega_{23}=0.005$ and $\Gamma/\Omega_{23}=0.025$. The three plots provide a suggestive view which resembles three photograms taken while a wave advances from right to left. This clearly illustrates that the parameter region corresponding to negative $\Delta$ is less affected by the presence of the decay and that a negatively higher value of $\Delta$ implies a greater robustness against dissipation. This fact can be understood in terms of the transfer mechanism process described in Sec.~\ref{sec:ideal}. When $\Delta$ is negatively large, nothing happens during the first and third crossings depicted in Fig.~\ref{fig:levelscheme}c and the complete transition occurs during the second crossing, which does not involve the state $\Ket{2}$. This means that the population of the decaying diabatic state $\Ket{2}$ is always zero or negligible, so that the system does not undergo any loss of probability. Of course, for this analysis to work, it is necessary that the coupling strength $\Omega_{13}$ is not negligible, otherwise no transition will occur around the second crossing (See appendix \ref{app:negdelta} for a more detailed analysis). 
On the contrary, for positively high values of $\Delta$ a $\Ket{1}\rightarrow\Ket{2}$ transition occur at the first crossing, followed by a $\Ket{2}\rightarrow\Ket{3}$ transition concomitant to the third crossing. Therefore, state $\Ket{2}$ is populated between the first and third crossing, which determines a loss of probability during the relevant time interval.

In Fig.~\ref{fig:Dissipation-2} it is plotted the efficiency of the population transfer in the presence of dissipation as a function of $\Delta$ and $\Gamma$, for different values of $\Omega_{13}$. From Fig.~\ref{fig:Dissipation-2}a we clearly see that for $\Omega_{13}=0$  there is no robustness for $\Delta<0$, while even a small $\Omega_{13}$ (Fig.~\ref{fig:Dissipation-2}b) is sufficient to have a higher efficiency. The situation improves for even higher values of $\Omega_{13}$, as in Fig.~\ref{fig:Dissipation-2}c.

It is worth mentioning that, as in the $\Delta=0$ case analyzed in Ref.~\cite{ref:Militello2019}, for very large decay rates and in the presence of a non negligible $\Omega_{13}$ there is a revival of efficiency due to a Hilbert space partitioning~\cite{ref:MilitelloFort2001,ref:Pascazio2002,ref:MilitelloPScr2011,ref:MilitelloQZE2011,ref:MilitelloQZE2012}.

\begin{widetext}

\begin{figure}[h]
\begin{tabular}{cccl}
\subfigure[]{\includegraphics[width=0.27\textwidth, angle=0]{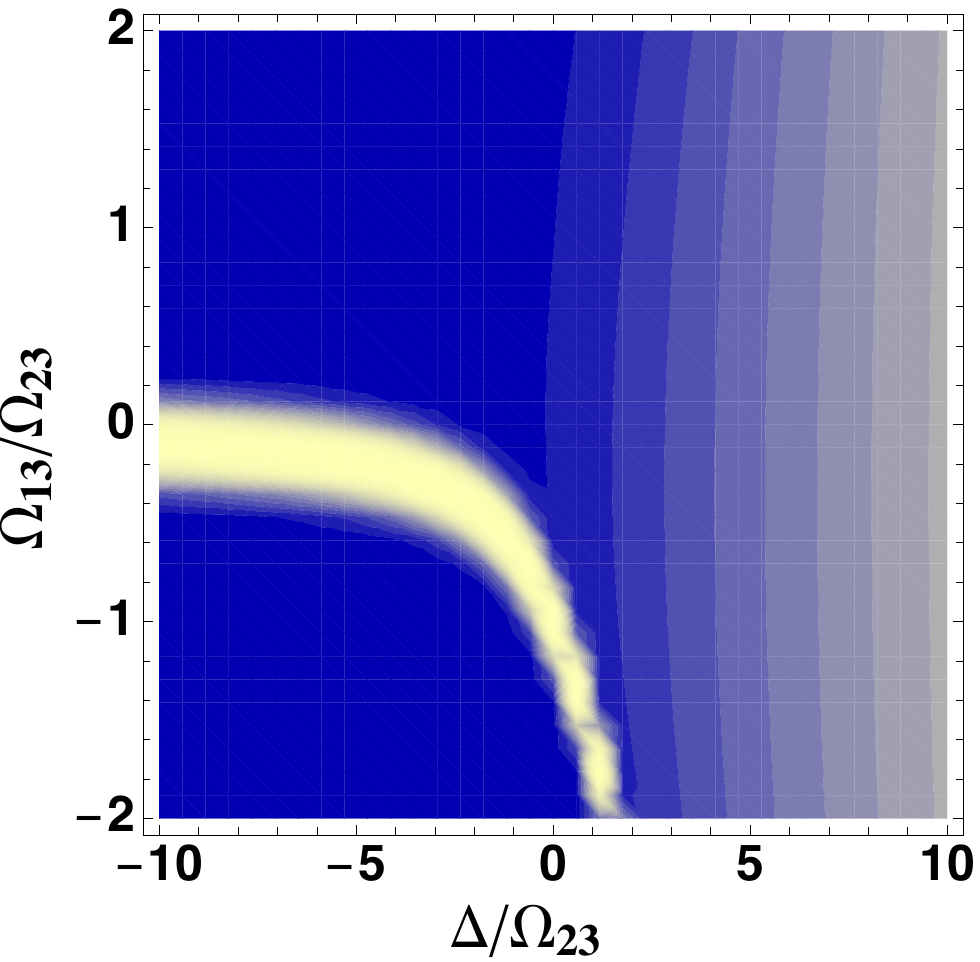}} & \qquad 
\subfigure[]{\includegraphics[width=0.27\textwidth, angle=0]{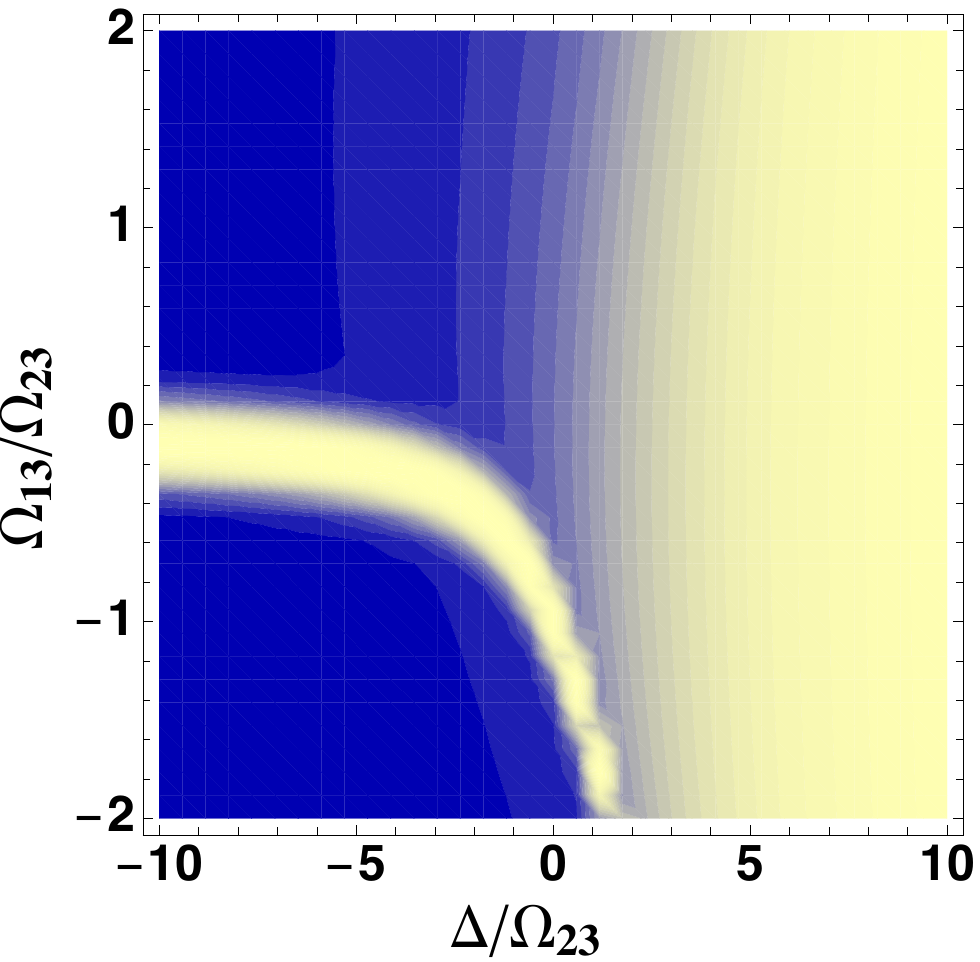}} & \qquad
\subfigure[]{\includegraphics[width=0.27\textwidth, angle=0]{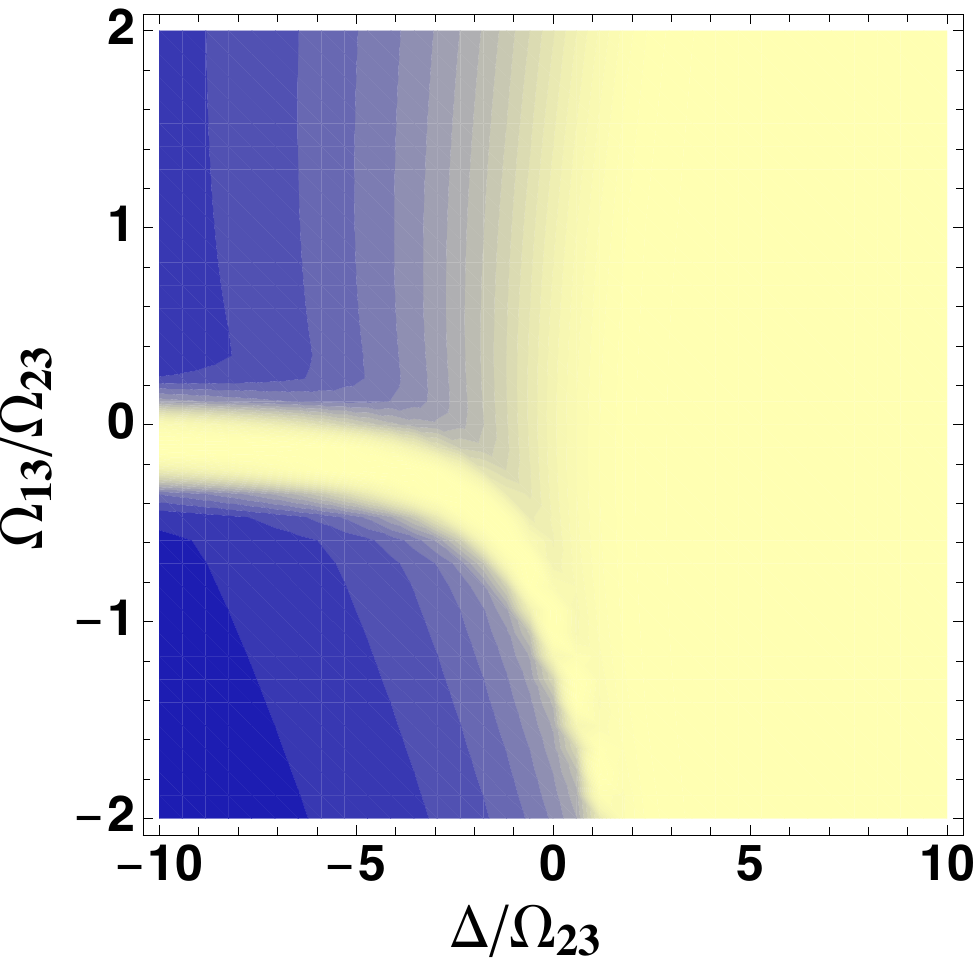}} &
\subfigure{\includegraphics[width=0.06\textwidth, angle=0]{fL1.pdf}} 
\end{tabular}
\caption{(Color online) Final population of state $\Ket{3}$ (obtained through a numerically exact resolution of the relevant Schr\"odinger equation) when the system starts in $\Ket{1}$ as a function of $\Omega_{13}$ and $\Delta$ (both in units of $\Omega_{23}$). The relevant parameters are: $\Omega_{12}/\Omega_{23}=1$, $\kappa/\Omega_{23}^2=0.1$, $\kappa T/\Omega_{23}=100$. Concerning the decay rate, the following values have been considered: $\Gamma/\Omega_{23} = 0.001$ (a), $\Gamma/\Omega_{23} = 0.005$ (b) and $\Gamma/\Omega_{23} = 0.025$  (c).} \label{fig:Dissipation-1}
\end{figure}

\begin{figure}[h]
\begin{tabular}{cccl}
\subfigure[]{\includegraphics[width=0.27\textwidth, angle=0]{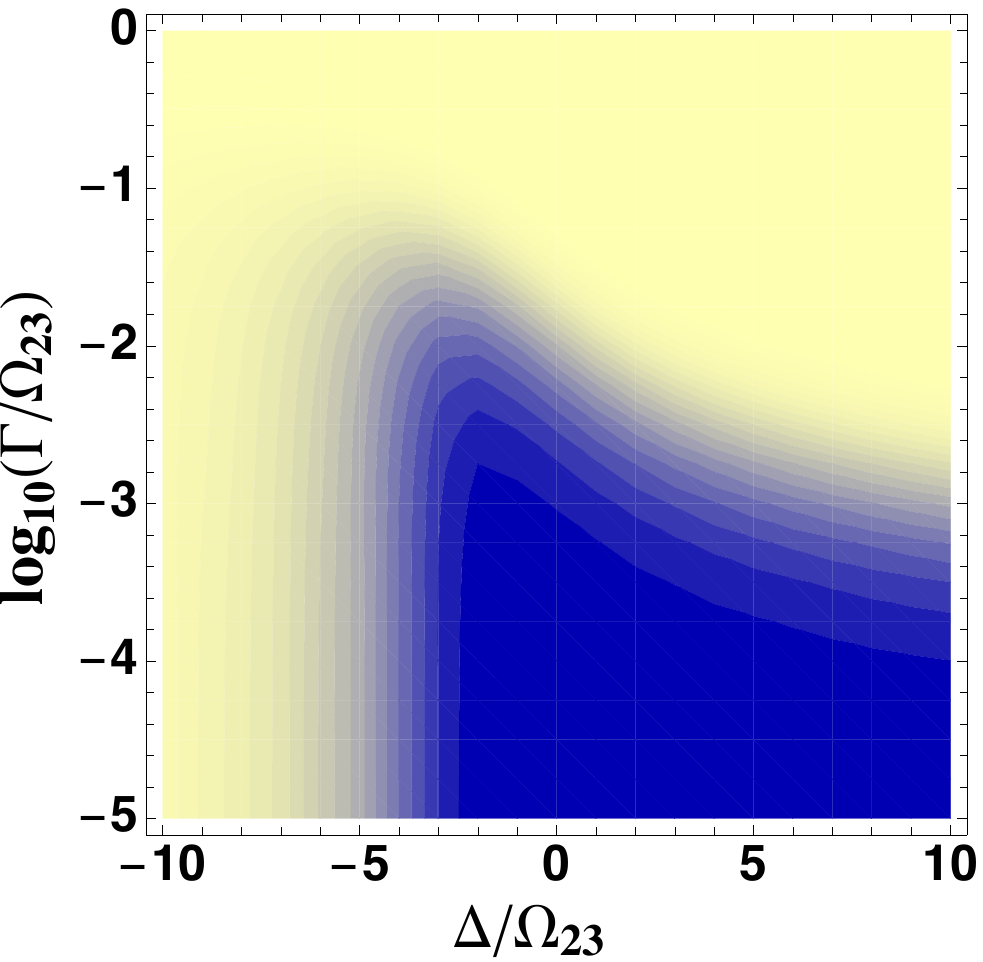}} & \qquad
\subfigure[]{\includegraphics[width=0.27\textwidth, angle=0]{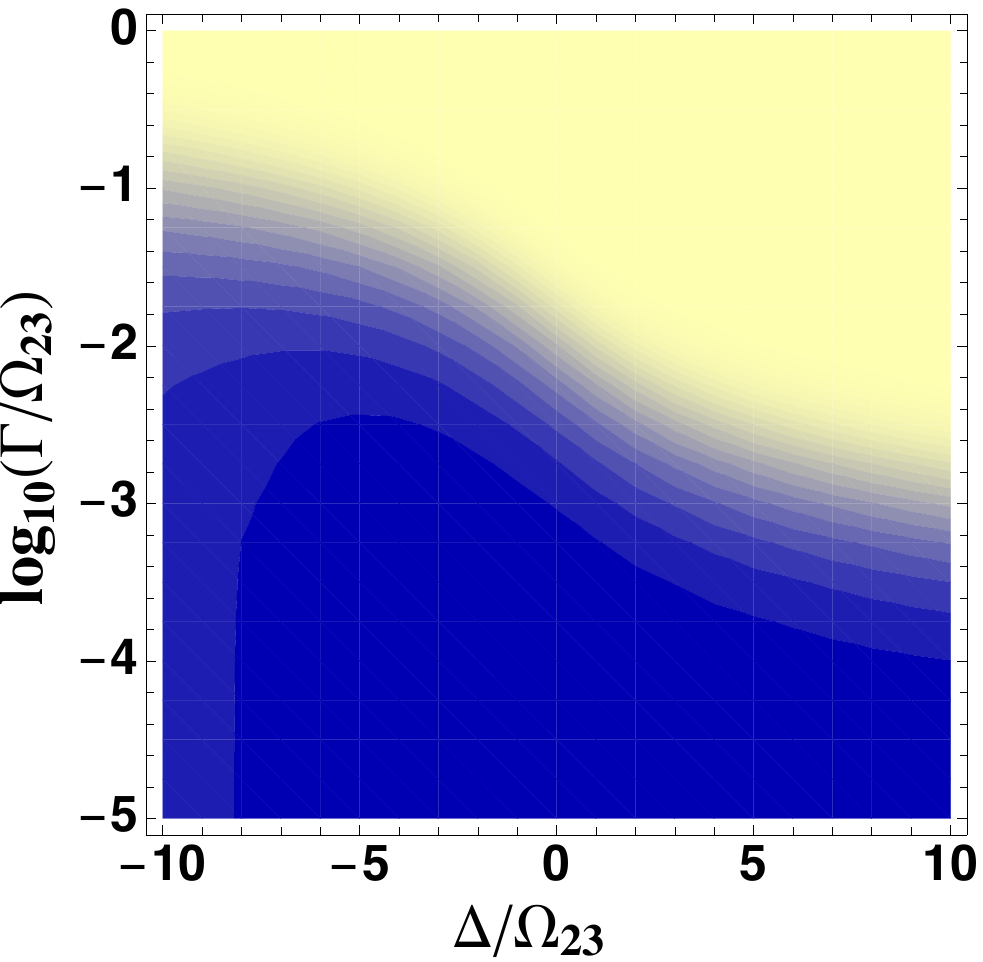}} & \qquad
\subfigure[]{\includegraphics[width=0.27\textwidth, angle=0]{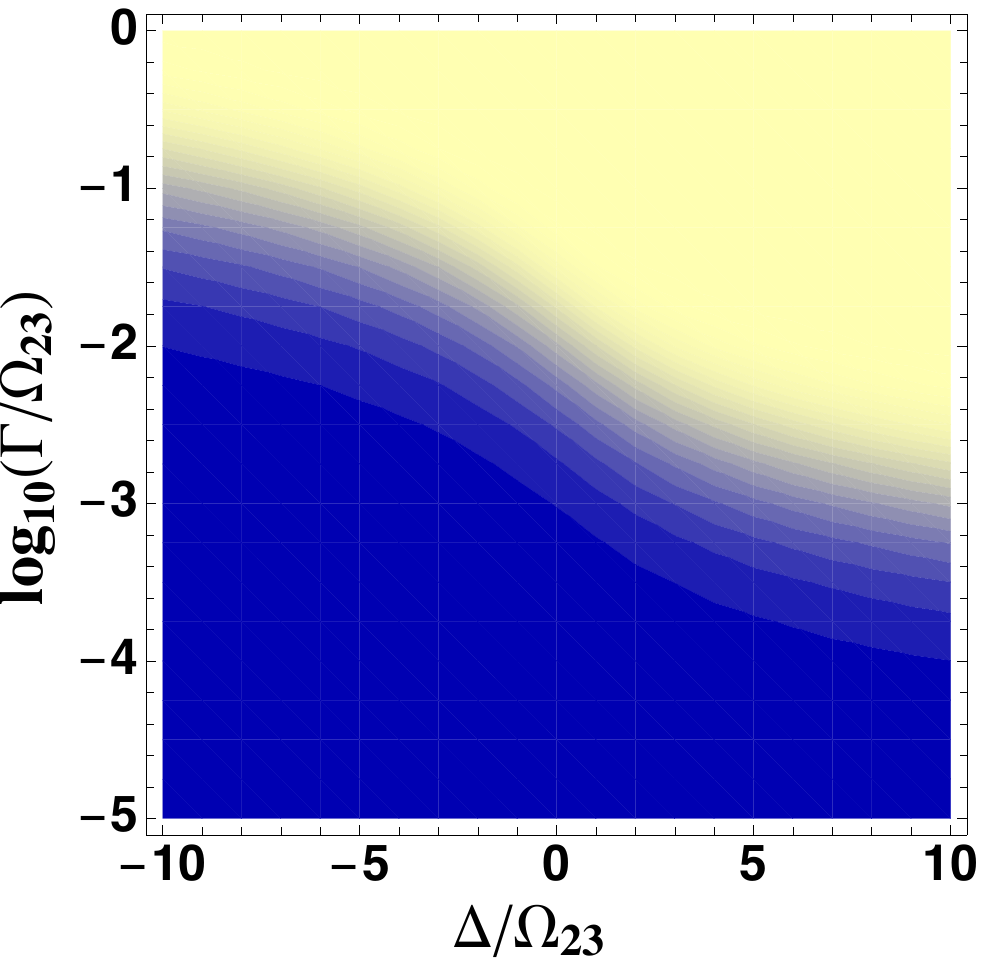}} &
\subfigure{\includegraphics[width=0.06\textwidth, angle=0]{fL1.pdf}} 
\end{tabular}
\caption{(Color online) Final population of state $\Ket{3}$ (obtained through a numerically exact resolution of the relevant Schr\"odinger equation) as a function of $\Delta$  (in units of $\Omega_{23}$) and $\Gamma$ (in units of $\Omega_{23}$ and in logarithmic scale), when the system starts being in the state $\Ket{1}$. The relevant parameters are $\Omega_{12}/\Omega_{23}=1$, $\kappa/\Omega^2_{23}=0.1$ and $\kappa T / \Omega_{23} = 100$ for the three figures. In (a) $\Omega_{13}=0$, in (b) $\Omega_{13}/\Omega_{23}=0.2$ and in (c) $\Omega_{13}/\Omega_{23}=0.5$.} \label{fig:Dissipation-2}
\end{figure}

\end{widetext}

\section{Conclusions}\label{sec:conclusions}

Summarizing, we have analyzed a three-state LZMS model characterized by the presence of a detuning (energy offset) in the state with static energy ($\Ket{2}$). We have shown that since the presence of the detuning changes the way the bare energies cross, it also influences the efficiency of the population transfer between the two states different from the detuned one.
In a different view, the detuning contributes in determining the minimum gap between the eigenvalues of the Hamiltonian, which is fundamental for establishing the validity of the adiabatic approximation and then the efficiency of the population transfer.  In particular, we have seen through several plots that a very low value of the gap, even at a single instant of time, is responsible for a dramatic diminishing of the efficiency.

We have also considered the effects of an interaction with the environment, which mainly has a negative influence on the transfer process. In the presence of decays involving the initial or target state of the process ($\Ket{1}$ and $\Ket{3}$, respectively) the diminishing of the efficiency is easily predicted as significant. If the decaying state is $\Ket{2}$, then the situation is more complicated, because, depending on the value of $\Delta$, such a state can be more or less involved in the dynamics. In the case of a limited (or negligible) involvement, the efficiency remains quite high in spite of the decaying process. This is the case for $\Delta<0$.

We conclude by commenting on what happens if we swap the roles of the initial and target states. In this case, the instantaneous eigenstate which must be used as a population carrier is the one corresponding to lowest energy, i.e., $\epsilon_3(t)$, according to the notation introduced before \eqref{eq:mingap}. The quantity that must be analyzed is then the minimun gap between the two lowest eigenvalues, which can be obtained as the minimum gap between the two highest eigenvalues of $-H(t)$, which is $H(t)$ associated to $-\kappa$, $-\Delta$ and  $\{-\Omega_{kj}\}$. Moreover, the minimum gap for $-H(t)$ is independent from the sign of $\kappa$, since the minimum is evaluated for $t$ (and then $\kappa t$) spanning symmetrically positive and negative values. Therefore, in order to check the validity of the adiabatic approximation when we want to have a $\Ket{3}\rightarrow\Ket{1}$ transition, we have to evaluate ${\cal G}$ for $\kappa$, $-\Delta$ and $\{-\Omega_{kj}\}$.

\appendix

\section{Analysis of the $\Delta<0$ dissipative case}\label{app:negdelta}

In this appendix we try to better support the independent crossing treatment and the consequent results in the $\Delta<0$ dissipative case.
Depending on the time instant, we can consider the Hamiltonian as the sum of different terms to be treated according to the perturbation theory, which will give us the possibility to predict the system behavior.
We assume a negatively large $\Delta$ ($\Delta<0$ and $|\Delta|\gg |\Omega_{ij}|$, $\forall i,j$), a non negligible $\Omega_{13}$ and a pretty small $\Gamma$. The state $\Ket{\psi(t)}=\sum_k c_k(t)\Ket{k}$ is assumed to start in the condition $a_1(-T)=1$, $a_2(-T)=a_3(-T)=0$.

For $|\kappa t| \gg |\Delta|$ (negatively large values of $t$) we have the diagonal terms as the \lq gross\rq\, part of the Hamiltonian and all the couplings as a perturbation: 
\begin{eqnarray}\label{appeq:a1}
\nonumber
H_D(t) = \left(
\begin{array}{ccc}
-\kappa t & 0 & 0 \\
0 & \Delta - \ii\Gamma & 0 \\
0 & 0& \kappa t  
\end{array}
\right)  + 
 \left(
\begin{array}{ccc}
0 & \Omega_{12} & \Omega_{13} \\
\Omega_{12} & 0 & \Omega_{23} \\
\Omega_{13} & \Omega_{23} &0
\end{array}
\right)\,. \\
\end{eqnarray}
The gross part leaves the initial state unchanged, while the perturbation induces small deviations. Introducing $\xi=(\max |\Omega_{ij}|)/|\Delta|$ ($\ll 1$), we can say that they are $o(\xi)$, getting $\Ket{\psi(t)} = \Ket{1} + o(\xi)$.

When it happens that $\kappa t \sim \Delta$ (negative values of $t$), the correct way the separate the terms is the following:
\begin{eqnarray}
\nonumber
H_D(t) &=& \left(
\begin{array}{ccc}
-\kappa t & 0 & 0 \\
0 & \Delta - \ii\Gamma & \Omega_{23} \\
0 & \Omega_{23} & \kappa t  
\end{array}
\right)  + 
 \left(
\begin{array}{ccc}
0 & \Omega_{12} & \Omega_{13} \\
\Omega_{12} & 0 & 0\\
\Omega_{13} & 0 & 0
\end{array}
\right)\,.\\
\end{eqnarray}
Also in this case, the state essentially given by $\Ket{\psi(t)}=\Ket{1}+o(\xi)$ is left unchanged, up to additional deviations $o(\xi)$. 
During the period when $|\Omega_{13}| \ll\kappa t \ll |\Delta|$ (positive and negative values around $t=0$) the appropriate separation is
\begin{eqnarray}
\nonumber
H_D(t) &=& \left(
\begin{array}{ccc}
-\kappa t & 0 &  \Omega_{13} \\
0 & \Delta - \ii\Gamma & 0  \\
 \Omega_{13} & 0 & \kappa t  
\end{array}
\right)  + 
 \left(
\begin{array}{ccc}
0 & \Omega_{12} & 0 \\
\Omega_{12} & 0 &  \Omega_{23}\\
0 &  \Omega_{23} & 0
\end{array}
\right)\,,\\
\end{eqnarray}
and a proper two-state Landau-Zener transition occurs up to terms of the order $o(\xi)$ due to the perturbation. Therefore, after this time interval one has $\Ket{\psi(t)}=(1-\ee^{-\pi\Omega_{13}^2/\kappa}) \Ket{3}+o(\xi)$. 

In the subsequent time interval where $-\kappa t \sim \Delta$ (positive values of $t$), the Hamiltonian can be rearranged as 
\begin{eqnarray}
\nonumber
H_D(t) &=& \left(
\begin{array}{ccc}
-\kappa t & \Omega_{12} & 0 \\
\Omega_{12} & \Delta - \ii\Gamma & 0 \\
0 & 0 & \kappa t  
\end{array}
\right)  + 
 \left(
\begin{array}{ccc}
0 & 0 & \Omega_{13} \\
0 & 0 &  \Omega_{23}\\
\Omega_{13} &  \Omega_{23} & 0
\end{array}
\right)\,,\\
\end{eqnarray}
which only slightly affects the state describing the system at this stage.
Finally, we consider $|\kappa t|\gg |\Delta|$ (positive values of $t$) and we get again the arrangement in \eqref{appeq:a1}, which essentially leaves the state unchanged. The net result is then roughly given by $\Ket{\psi(T)} = (1-\ee^{-\pi\Omega_{13}^2/\kappa}) \Ket{3}+o(\xi)$.

\end{document}